\documentclass[journal]{IEEEtran}

\usepackage{amsmath,color,url,mathrsfs,upgreek,amsthm,enumerate,multirow}
\usepackage{hyperref,mathrsfs,subcaption,graphicx,amsfonts,algpseudocode}
\usepackage[linesnumbered,algoruled,boxed,lined]{algorithm2e}
\usepackage[normalem]{ulem}

\newtheorem*{theorem*}{Theorem}

\newcommand{\bpg}{\begin{paragraph}{}}
\newcommand{\epg}{\end{paragraph}}
\newcommand{\bit}{\begin{itemize}}
\newcommand{\eit}{\end{itemize}}
\newcommand{\beq}{\begin{equation}}
\newcommand{\eeq}{\end{equation}}
\newcommand{\beqn}{\begin{equation*}}
\newcommand{\eeqn}{\end{equation*}}
\newcommand{\beqa}{\begin{equation}\begin{aligned}}
\newcommand{\eeqa}{\end{aligned}\end{equation}}
\newcommand{\beqna}{\begin{equation*}\begin{aligned}}
\newcommand{\eeqna}{\end{aligned}\end{equation*}}

\newcommand{\enum}{\begin{enumerate}}
\newcommand{\enuma}{\begin{enumerate}[(a)]}
\newcommand{\eenum}{\end{enumerate}}
\newcommand{\norm}[1]{||#1||}

\newcommand{\inprod}[1]{\langle #1 \rangle}

\newcommand{\lp}{\left(}
\newcommand{\rp}{\right)}

\newcommand{\abs}[1]{\left|#1\right|}

\newcommand{\bolx}{\boldsymbol{x}}

\newcommand{\boll}{\boldsymbol{\lambda}}
\newcommand{\bell}{\boldsymbol{\ell}}

\newcommand{\bol}[1]{\boldsymbol{#1}}

\newcommand{\AG}[1]{{\color{black}#1}}

\definecolor{dkgreen}{rgb}{0,0.6,0}
\definecolor{gray}{rgb}{0.5,0.5,0.5}
\definecolor{mauve}{rgb}{0.58,0,0.82}

\begin{document}
\title{Thresholded Non-Uniform Fourier Frame-Based Reconstruction for Stripmap SAR}

\author{John~McKay,~
        Anne~Gelb,~%
        Suren~Jayasuriya,~
       Vishal~Monga~
\thanks{McKay is affiliated with the Applied Research Laboratory \& Monga is affiliated with the Dept. of Electrical Engineering both under the Pennsylvania State University. E-mail:John.McKay@psu.edu}
\thanks{Gelb is affiliated the Dept of Math at Dartmouth College.}
\thanks{Jayasuriya holds joint appointments between the departments of Arts, Media and Engineering as well as Electrical, Computer, and Energy Engineering both of Arizona State University.}
\thanks{McKay, Jayasuriya, and Monga are hold IEEE memberships.}
\thanks{Funding Support by ONR N00014-15-1-2042, AFOSR FA9550-18-1-0316}}

\maketitle

\begin{abstract}
Fourier domain methods are fast algorithms for SAR imaging. They typically involve an interpolation in the frequency domain to re-grid non-uniform data so inverse fast Fourier transforms can be performed. In this paper, we apply a frame reconstruction algorithm, extending the non-uniform fast Fourier transform, to stripmap SAR data. Further, we present an improved thresholded frame reconstruction algorithm for robust performance and improved computational efficiency. We demonstrate compelling results on real stripmap SAR data.
\end{abstract}


\IEEEpeerreviewmaketitle

\section{Introduction}
\IEEEPARstart{F}{ourier} domain methods (FDMs) are the preferred approach for computationally efficient SAR imaging~\cite{soumekh1999synthetic}. Backprojection and sparsity-based compression methods yield high quality image but at computational costs far exceeding FDMs~\cite{sommer2015comparison, potter2010sparsity}.  FDMs, however, are limited by a requirement to re-grid the acquired non-uniformly spaced data so an inverse fast Fourier transform (IFFT) can be applied. These interpolations incur global errors that degrade subsequent SAR image reconstruction quality for FDMs.

Non-uniform fast Fourier transforms (NUFFTs) are efficient, accurate alternatives to standard FDMs, particularly for range-migration algorithms~\cite{andersson2012fast,cheema2014power, fan2014polar}. NUFFTs do not suffer from beam-width limitations as do chirp-z transforms~\cite{li2014interpolation} and its computational complexity is $\mathcal{O}(N^2\log N)$ for $N$ data points. However the reconstruction quality of NUFFTs is still sub-par. Quadrature weights utilized in NUFFTs cause limited resolution, especially with reduced sampling~\cite{andersson2012fast,gelb2014frame}. 

This letter extends the robustness of NUFFTs to work on \textbf{real} SAR data, improving reconstruction quality while maintaining near-equivalent computational efficiency. We utilize recent theoretical work with frame methods which generalize the underlying NUFFT algorithm~\cite{gelb2014frame}. We will refer to this method as the non-uniform Fourier frame-based reconstruction, which we abbreviate (FFR). This theory represents the non-uniform Fourier data with a redundant set of basis functions, and derives a better weighting scheme for the NUFFT. While mathematically elegant, it has not been applied to real data for inverse imaging problems like MRI, SAR, or sonar. 

In this letter, we present the first application of FFR to stripmap SAR data. In addition, we introduce thresholded non-uniform Fourier frame-based reconstruction (tFFR) to better address the limitations of FFR including faster processing. This provides a compelling alternative to current interpolation methods and NUFFTs while preserving computational efficiency. Our contributions are the following:
\begin{enumerate}
\item The first application of FFR for real-world synthetic aperture imaging.
\item A novel thresholding algorithm, tFFR, to improve the FFR in both reconstruction quality and processing speed.
\item Validation of tFFR on actual stripmap SAR data, achieving better performance than the NUFFT, the original FFR, and Stolt interpolation baselines, while maintaining comparable computational efficiency.
\end{enumerate}


\section{Background}
\label{sec:background}
\subsection{Stripmap SAR Processing}\label{sec:sar}
\begin{figure}\centering
\includegraphics[width=.9\columnwidth]{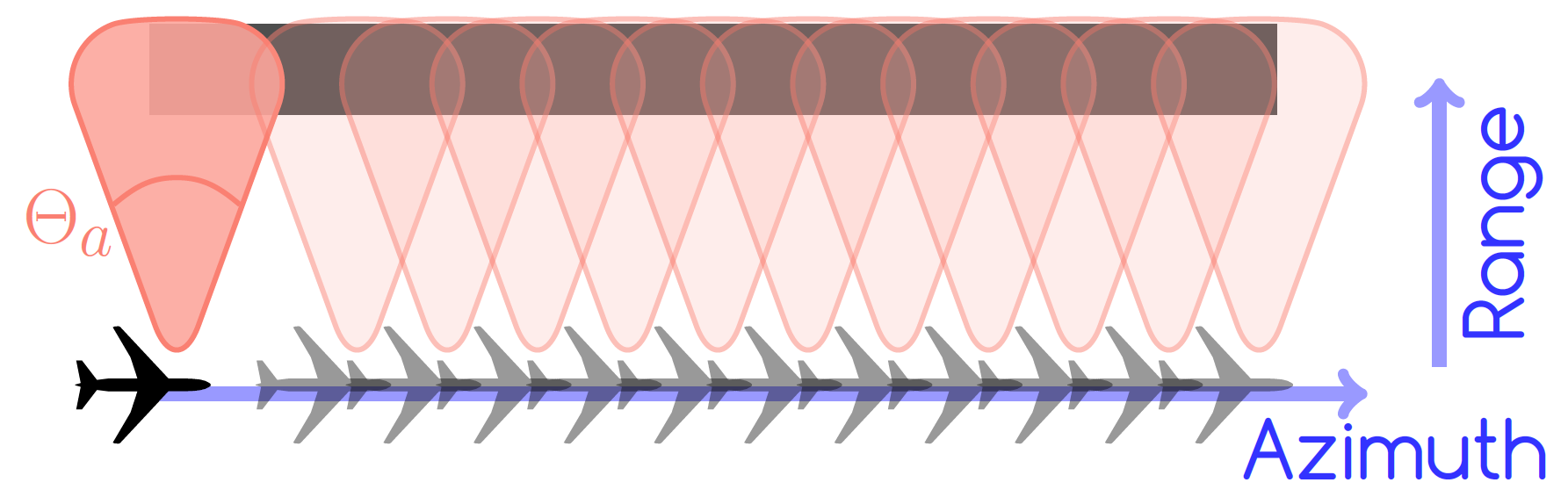}
\caption{Depiction of stripmap SAR. A vehicle travels a straight path along the azimuthal direction and emits a signal with beamwidth $\Theta_a$ in the perpendicular range direction.}\label{fig:stripmap}
\end{figure}

Stripmap SAR is the imaging technique where a vehicle equipped with the appropriate SAR antenna(s) and receiver(s) travels in a straight path, capturing the perpendicular scene (see Figure~\ref{fig:stripmap}). Stripmap SAR maximizes the area surveyed and has the finest cross-range resolution possible without interrupted coverage~\cite{melvin2010principles}. Collected data are typically processed through a range migration algorithm (RMA) also known as ``$\omega$-k'' processing~\cite{cumming2003interpretations}. This maps signals to the frequency domain, performs match-filtering to perform synthetic aperture refocusing, and then uses a Stolt interpolation to reconstruct the image from the received waveform~\cite{melvin2010principles,vandewal2007efficient}.

\begin{figure}\centering
\includegraphics[width=.8\columnwidth]{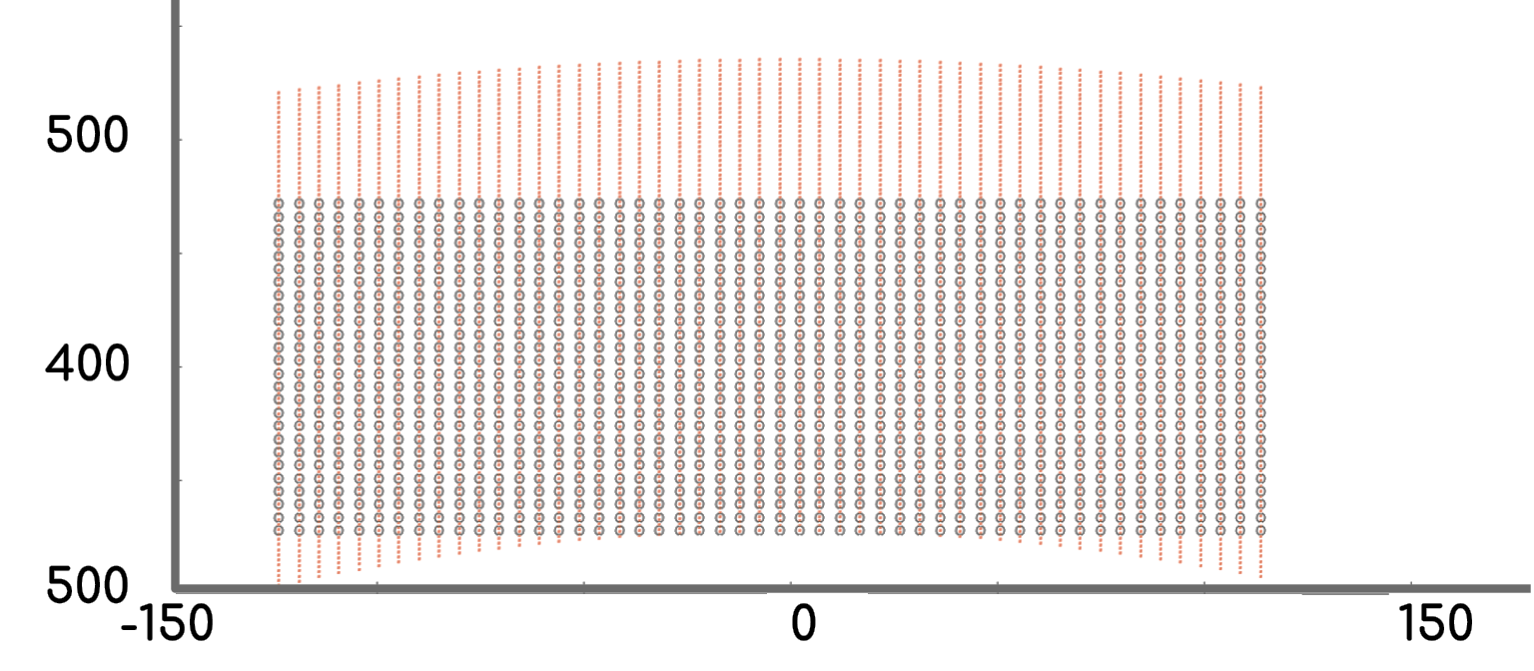}
\caption{Fourier sampling arrangement $\boldsymbol{\lambda}_n$ (orange) and desired grid $\ell_m$ (black) for "Go State" image (see Section 4).}\label{fig:sampling}
\end{figure}

Before Stolt interpolation, the radar data are non-uniform in the range while uniform in the azimuthal direction. Stolt interpolation aligns non-uniform frequency points onto a grid such that an efficient $2D$ IFFT can be used~\cite{vandewal2007efficient}. Figure~\ref{fig:sampling} depicts an example stripmap SAR data arrangement in the Fourier domain as well as a desirable band-limited grid for the data to be interpolated upon. This paper seeks to improve the reconstruction quality beyond Stolt using frame-theoretic methods.

\subsection{Non-Uniform Fast Fourier Transforms}\label{sec:fcg}

An alternative to Stolt interpolation is the Non-uniform Fast Fourier Transform (NUFFT), which achieves better image reconstruction at slightly higher computational cost \cite{andersson2012fast}. To formulate the NUFFT, first consider the underlying inverse imaging problem: Let $f(\boldsymbol{x})$ be the complex valued radar image we wish to reconstruct, $\hat f(\boldsymbol{\lambda}_n)$ its Fourier coefficients with $\{\boldsymbol{\lambda}_n=(\lambda^{1}_n,\lambda^{2}_n)\}_{n=1}^N$ the $2D$ frequency domain coordinates with a non-uniform arrangement given by stripmap SAR as shown in Figure~\ref{fig:sampling}. The image is reconstructed from the discrete approximation of the inverse Fourier transform:

\beqa\label{eq:fstraight}
f(\bolx) {\approx} \sum_{n=1}^N\alpha_n\hat{f}(\boll_n)e^{2\pi i \boll_n^T\bolx}.
\eeqa

If $\{\boldsymbol{\lambda}_n\}$ lied on a uniformly spaced grid, Equation~\eqref{eq:fstraight} becomes an inverse discrete Fourier transform (IDFT) with $\alpha_n=\frac{1}{N}$. When the data are not equally spaced, $\alpha_n \ne \frac{1}{N}$ and must be chosen carefully to ensure an accurate approximation and the IDFT cannot be directly applied.

To solve this problem, NUFFTs measure the deviation of the received frequencies from a uniformly spaced grid, and then weights their contributions in the inverse transform using a smooth window function~\cite{andersson2012fast}. Formally, we write our original radar image multiplied by this window as $g = fw$, and our approximation becomes the following:
\beqa
\hat g(\xi) \approx \sum\limits_{n=1}^N \alpha_n\hat f(\boldsymbol{\lambda}_n)\hat w(\xi-\boldsymbol{\lambda}_n)e^{2\pi i\boldsymbol{\lambda}_n^T\xi}.
\eeqa
where we utilized the Fourier convolution property. Typically, $w$ is a Gaussian which yields a Gaussian $\hat w$~\cite{greengard2004accelerating}.

We can control the weighting of $\lambda_n$ frequencies relative to an uniform grid of points $\{\bell_m\}_{m=1}^M$ (like in Figure \ref{fig:sampling}) by setting $\xi=\bell_m$. It follows that $g$ is then:
\beqa\label{eq:g}
g(\bolx) \approx \sum\limits_{n=1}^N\sum\limits_{m=1}^M\alpha_n\hat{f}(\boll_n)\hat{w}(\bell_m-\boll_n)e^{2\pi i \bell_m^T\bolx}.
\eeqa

\begin{figure}\centering
\includegraphics[width=\columnwidth]{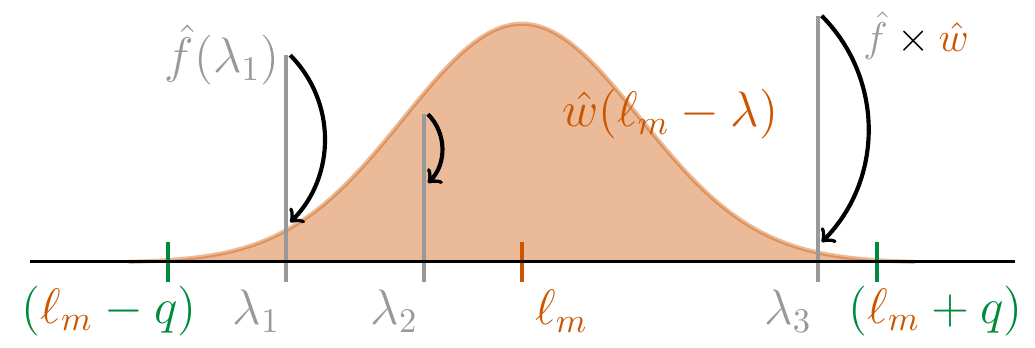}
\caption{Illustration of NUFFT; the further Fourier samples $\hat f(\lambda_n)$ are from $\ell_m$, the more they are mollified by $\hat w(\ell_m-\lambda_n)$.}\label{fig:nufft}
\end{figure}

The radar image is recovered as $f=\frac{g}{w}$. To improve speed, only $\boll_n$ within a chosen bound $q$ are considered for each $\ell_m$. An illustration of the NUFFT idea is given in Figure \ref{fig:nufft}. We define the \textbf{NUFFT} formula for the approximation of $f$:
\beqa\label{eq:Anu}
f_{NUFFT}(\bolx) := \sum\limits_{m=1}^M\sum\limits_{\norm{\boll_n-\bell_m}<q}\alpha_n\hat{f}(\boll_n)\hat{w}(\bell_m-\boll_n)\tfrac{e^{2\pi i \bell_m^T\bolx}}{w(\bolx)}.
\eeqa

A key to the success of NUFFTs is the choice of weights $\alpha_n$. A quadrature scheme using trapezoidal rule usually performs well in SAR settings as the non-uniform data are densely sampled~\cite{andersson2012fast,fan2014polar}. We note that in MRI, where $\alpha_n$ are known as density compensation factors, iterative methods are often used to calculate $a_n$ in more sparse settings~\cite{pipe1999sampling}. In the next section, we present methods using frame theory to derive better estimates for $\alpha_n$. In particular, we present our novel thresholding method to regularize the choice of $\alpha_n$ and achieve high quality and cost efficient image reconstructions. 

\section{Thresholded Non-Uniform Fourier Frame-Based Reconstruction}
NUFFTs, while computationally efficient, are sensitive to the underlying non-uniform sampling pattern. This results in low quality image reconstructions if $\alpha_n$ are not chosen carefully~\cite{viswanathan2010reconstruction}. This section demonstrates how non-uniform Fourier frame-based reconstruction (FFR) addresses these issues inherent to NUFFTs. 
\AG{Below we demonstrate that the FFR method can be further improved for stripmap SAR applications.  In particular}
we introduce thresholded non-uniform Fourier frame-based reconstruction, which we abbreviate as tFFR, as a novel, robust method achieving state-of-the-art performance on real-world SAR data, while not sacrificing computational efficiency. In Section IV we will present extensive experimental results that validate these claims. 

\subsection{Non-uniform Fourier Frame-based Reconstruction (FFR)}

FFR was first proposed in \cite{gelb2014frame}. The method interprets non-uniform Fourier data as samples of a Fourier \emph{frame}.  A frame is a set of vectors that span a vector space with no linear independence requirement and is often viewed as an overcomplete basis~\cite{christensen2008frames}. This is useful as real-world data often fail to form a basis for a truncated Fourier approximation, but can be used with frames. FFR determines the weights $\alpha_n$ in Equation \eqref{eq:Anu} to improve the robustness of the NUFFT~\cite{gelb2014frame}.

\AG{FFR works by exploiting the so-called admissible frame approxiation (FA) method \cite{song2013approximating}. Using the notation from Section II, consider the  image $f(\boldsymbol{x})$, sampled non-uniform frequencies $\lambda_n$, window function $w$ of the NUFFT, and the uniform grid of frequencies, $\{\bell_m\}_{m = 1}^M$.}
The \textbf{FA}~\footnote{This is \AG{an example of} a Fourier FA. Different admissible frames can be \AG{constructed} depending on the application \cite{song2013approximating}.} is given as~\cite{song2013approximating}
\beqa\label{eq:Afa}
f_{FA}(\bolx) \approx \sum\limits_{m=1}^M\sum\limits_{n=1}^N b_{m,n}\hat f(\boll_n)\frac{e^{2\pi i \bell_m^T\bolx}}{w(\bolx)},
\eeqa
where the matrix of coefficients $B = [b_{m,n}]$ is defined by the Moore Penrose pseudo-inverse $B=\Psi^\dagger$ where $\Psi$ is defined as
\beqa
\label{eq:Psi}
\Psi = [\inprod{e^{2\pi i \boll_n^T\bolx},\frac{e^{2\pi i \bell_m^T\bolx}}{w(\bolx)}}].
\eeqa
\AG{Observe that (\ref{eq:Afa})  provides a robust way to compute the NUFFT, \cite{song2013approximating}. Standard NUFFT (where the quadrature coefficients (\ref{eq:Anu}) are heuristically chosen) can sometimes lack robustness, especially for highly non-unfirom  sampling \cite{viswanathan2010reconstruction}.} However, in the the form given, (\ref{eq:Afa})  requires $\mathcal{O}(NM^2)$ complexity from the pseudo inverse calculation, making it impractical for many FDM settings.

{An efficient implementation for (\ref{eq:Afa}) was constructed in \cite{gelb2014frame}} and, with an assumption that $\hat w$ is negligible outside a small window, it was shown that 
\beqa\label{eq:bound}
\norm{f_{NUFFT}-f_{FA}}_F^2 \leq K\norm{W\Psi D - I}_F^2,
\eeqa
where $W = [\hat w(\bell_m-\boll_n)]$ is the matrix of window function values, $K$ is a constant, and $D$ is a diagonal matrix with $D_{nn} = \alpha_n$. The bound in  Equation \eqref{eq:bound} is exploited to generate the {\em non-uniform Fourier frame-based reconstruction} (FFR) which inherits the accuracy of (\ref{eq:Afa}) at only a slightly higher computational cost of the standard NUFFT~\cite{gelb2014frame,song2013approximating}. The approximation for $f$ is:
\beqa\label{eq:notDiagonal}
f_{FFR}(\bolx) = \sum\limits_{m=1}^M\sum\limits_{n=1}^N D_{n,m} \hat f(\boll_n) \hat w(\bell_m-\boll_n)\tfrac{e^{2\pi i\bell_m^T \bolx}}{w(\bolx)}
\eeqa
where $D_{n,m} = B_r((W\Psi)^\dagger_{n,m})$ with $B_r(X_{n,m})=X_{n,m}$ if $|m-n| \leq r$ and zero otherwise. In other words, $D$ is a banded diagonal matrix. The matrices $W$ and $\Psi$ are designed to be localized along the diagonal, so a band captures most information~\cite{song2013approximating}. The size of the band theoretically determines a trade-off between quality and cost: larger bands better approximate $f_{FA}(\bolx)$, while smaller bands approximate  $f_{NUFFT}(\bolx)$\footnote{The coefficients for the NUFFT derived from \eqref{eq:bound} are determined by minimizing the right hand side of (\ref{eq:bound}) with $D = diag(\alpha_1,\cdots,\alpha_N)$.}. 

{In real imaging systems, FFR with $r>1$ may perform poorly due to noise or limited sampling.  In Section IV we propose a thresholding strategy to alleviate these issues for the approximation.} 

\subsection{Thresholding for Improved FFR Performance}
In the stripmap SAR setting, the data lie close to the uniform grid so $D$ can be well approximated by a truncated diagonal matrix.  Moreover, data points collected away from the equal-spaced grid are likely corrupted by noise and populate the off-diagonal entries of $\Psi$. We mitigate these effects by defining a {\em threshold} on the matrix $D$. We want to threshold $D$ with $D_{nm}= R_\uptau((W \Psi)^{\dagger}_{nm})$ where $R_\uptau(x) = x \text{ if } x \geq \tau$ and zero elsewhere. This can be interpreted as preserving only the most correlated values of the frame approximation coefficients. Again, the way $W \Psi$ is structured, such values will lie upon or near the diagonal~\cite{gelb2014frame}.  Equation~\eqref{eq:notDiagonal} with such $D$ is our \emph{thresholded FFR} algorithm (tFFR).

\begin{algorithm}[t]
\DontPrintSemicolon
\KwIn{
Fourier data: $\{\hat f((\lambda_n^1, \lambda_p^2)\}_{n=1,p=1}^{N,P}$; grid of frequencies:
$\{\ell_m\}_{m=1}^M$, threshold $\uptau>0$
}
\KwOut{SAR image $f$}
Initialize $\hat f_{\text{new}} = \boldsymbol{0}$\;;
\For{n = 1:$N$} {
	$\bol{\hat{f}_p}\gets \text{vec}(\{\hat I(\lambda^1_n,\lambda^2_p)\}_{p=1}^{P})$\;
	$W\gets [\hat{w}(\ell_m-\lambda^2_p)]_{m,p}$\;
	$\Psi_{p,m} = [\inprod{\exp(2\pi i \lambda^2_p x),\exp(2\pi i\ell_m x)/w(x)}]_{p,m}$\;
	$D\gets (R_\tau((W\Psi)^\dagger)$\;
	$\bolx\gets\tfrac{1}{M}[1,\dots,M]^T$\;
	$\hat f_{\text{new}}(\lambda^1_n,:)\gets W D\bol{\hat{f}}\oslash w(\bolx)$
}
$f\gets \text{$2D$\_IFFT}(\hat \hat f_{\text{new}})$\;
\Return{$f$}\;
\caption{Stripmap SAR Imaging with tFFR}
\label{alg:rma}
\end{algorithm}

To analytically justify our threshold, we derive an additional bound to Equation \eqref{eq:bound} to illustrate a fundamental trade-off in the structure of $D$:
\beqa\label{eq:rewrite}
\norm{W\Psi D - I}_F^2 \leq \norm{D}_F^2\norm{W\Psi - D^\dagger}_F^2
\eeqa
The error derived in Equation \eqref{eq:bound} is related to either how fully or sparsely populated $D$ is. If it is fully populated, then $\norm{W\Psi D-I}$ is minimized. Otherwise, if $D$ is sparse, then $\norm{D}_F^2$ is minimized. 
As stated previously, the off diagonal entries of $W\Psi$ are susceptible to noise in real data. Thus, we usually use a high threshold $\tau$, which yields an extremely sparse $D$ in practice and improved computational complexity.

\noindent \textbf{Computational Complexity:} As previously noted, we expect $W\Psi$ to be diagonally dominant. Thus, $\uptau$ can be chosen large enough so only diagonal elements survive. Isolating the diagonal of $W\Psi$ to calculate $D$ drastically reduces the computational burden as now $\mathcal{O}(N)$ operations are required to calculate $W\Psi$ and $c<N$ are used because of our threshold. This is far less than the $\mathcal{O}(N\log N)$ required for FFR. Once $D$ is calculated, both FFR and tFFR finish their reconstructions with IFFTs requiring $\mathcal{O}(M\log M)$ operations. 

We summarize the computational complexity of the three frame methods: (1) FA requires $\mathcal{O}(NM^2)$ operations, (2) FFR requires $\mathcal{O}(N^2\log N + M\log M)$, and (3) tFFR requires $\mathcal{O}(N + M\log M)$. In practice, we find tFFR performs 5 times faster than FFR on our real experiments in Section IV.

\subsection{tFFR Algorithm for Stripmap SAR Imaging}\label{subsec:discrete}

To implement tFFR for stripmap SAR, we first observe that the $2D$ radar data can be viewed as a set of $1D$ slices along the azimuthal direction. Because of this, we reconsider our Fourier data as $\{\hat f((\lambda_n^1,\lambda_p^2))\}_{n,p=1}^{N,P}$ with $\lambda_n$ the azmuthal direction. In this way, we can focus our efforts on the associated non-uniformly sampled range data $\lambda_p$. The non-uniformity is addressed by calculating $\Psi$, $W$ and then finding $D = R_\tau((W \Psi)^\dagger)$. For stripmap settings, it suffices to just calculate  the diagonal of $W\Psi$ for efficiency. Importantly, $D$ only has to be calculated for each slice and the non-uniformity is ``corrected'' via the product $W D \bol{\hat f}$ where $\bol{f}$ is the vectorized slice.  Once this and the division of the window function is done for each azimuthal sample, the entire set is uniformly distributed and can be inverted via a $2D$ IFFT, yielding the image $I$.  The process is detailed in Algorithm~\ref{alg:rma}.  Note that the formula for $\Psi$ is found analytically.

\section{Experiments}\label{sec:experiments}

\begin{figure}\centering
\includegraphics[width=\columnwidth]{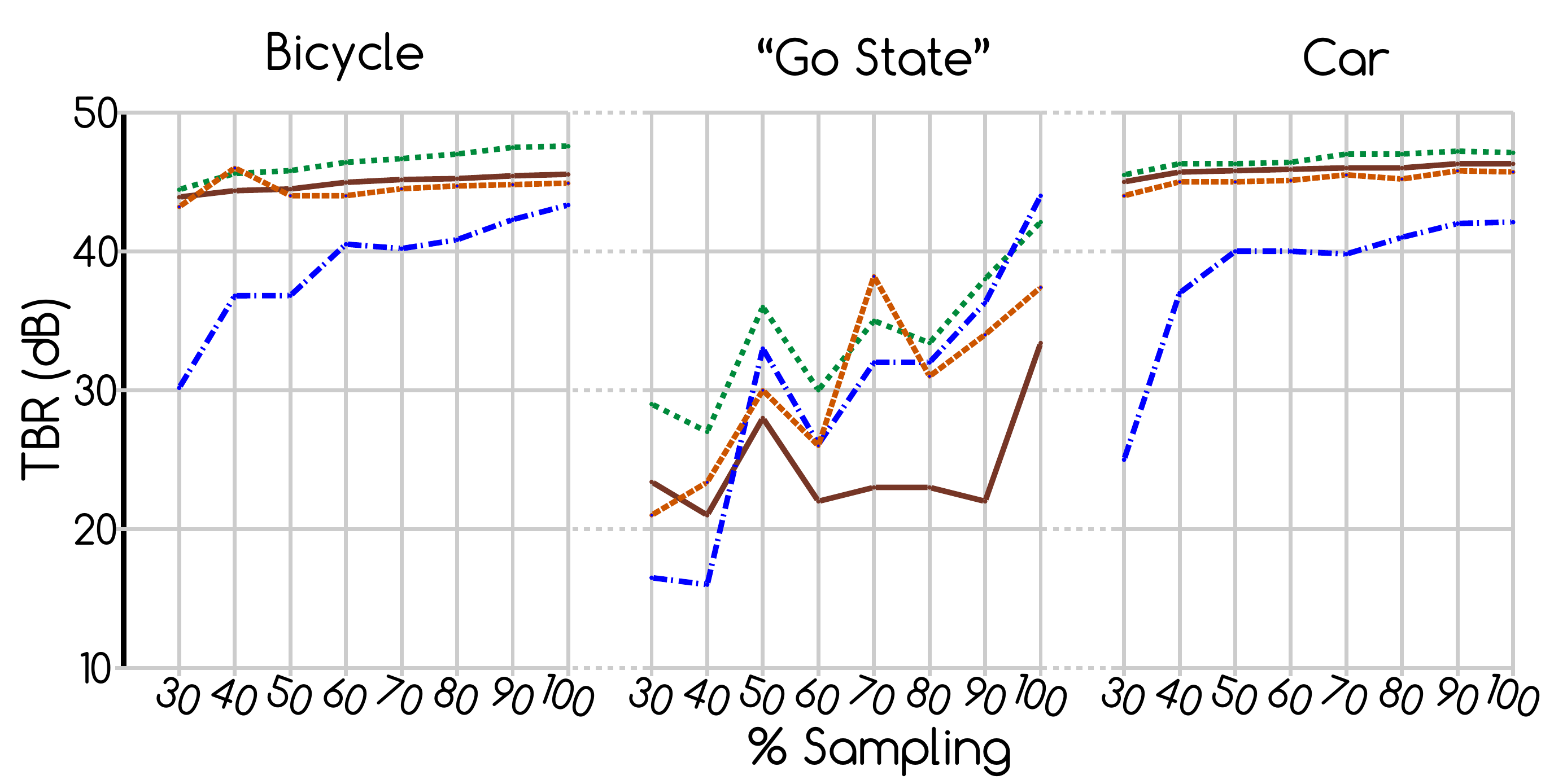}
\includegraphics[width=\columnwidth]{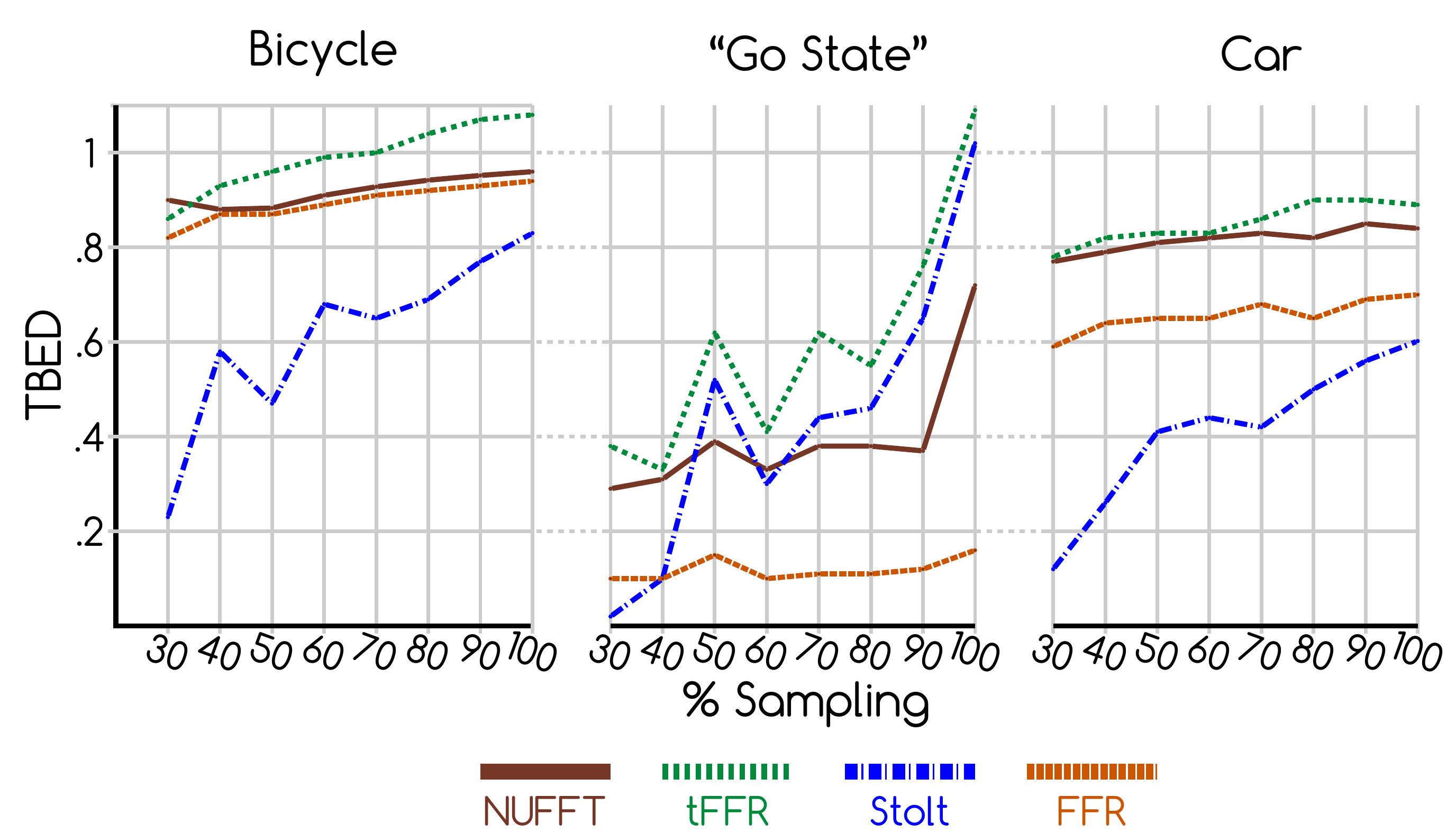}
\caption{TBR (top) and TBED (bottom) values for each method and imaging scenario.}\label{fig:tbr_tbed}
\end{figure}

In this section, we validate our proposed tFFR algorithm on real-world stripmap SAR data. We compare against the performance of baselines such as Stolt interpolation, NUFFT, and original FFR. \textbf{Note:} This is the first application of NUFFTs and FFRs to real-world strimap SAR data to the best of our knowledge. We show that our improved thresholding allows for robust reconstructions even in limited sampling settings. We propose a series of metrics to quantitatively evaluate our results and measure the computational speed. 

\textbf{Data:} We use data provided in~\cite{charvat2006low}, where a cost efficient X-band rail SAR system was designed to image three objects: a bicycle, a side view of a car, and a push-pin arrangement spelling ``Go State''. Similar to what was done in~\cite{samadi2011sparse}, we test our algorithms over stratified sampling cases to get an idea of how they do when frequencies are restricted, e.g. as a result of sensor limitations or jamming. We refer to the full set of data as 100\% and subsequent down-sampled cases are restrictions to 90\%, 80\%, etc.

\textbf{Baselines:} We implement three baselines: (i) Stolt interpolation, (ii) NUFFT, and (iii) FFR. For the NUFFT, we use the same window function with trapezoidal-rule-weights for $D$ as previous work~\cite{subiza2004nonuniform,greengard2004accelerating,andersson2012fast}. For the FFR without thresholding, we set $D$ with a band of $r=2$ as proposed in~\cite{gelb2014frame}. Following the experiments in~\cite{gelb2014frame}, we use the window function $w(x)=e^{-0.01\abs{x-.5}}$ for the NUFFT, FFR, and our tFFR algorithm. For our tFFR method, we let $\uptau=0.97$.

\textbf{Metrics:} While qualitatively viewing the images can yield some insight into reconstruction quality, we require quantitative metrics, especially since there is no ground truth image for comparison. As in  \cite{samadi2011sparse}, we utilize two reference-less measures: the  target-to-background ratio (TBR) \cite{cetin2003feature,benitz1997high} and target-to-background entropy difference (TBED) \cite{clark1991image}, respectively given by
\beqa\label{eq:metrics}
\text{TBR} &= 20\log_{10}\lp \tfrac{\max_{i\in\mathcal{T}}(\mathcal{I}_i)}{\text{mean}_{i\in\mathcal{B}}(\mathcal{I}_i)}\rp\\
\text{TBED}&=\abs{\text{entropy}_{i\in\mathcal{T}}(\mathcal{I})-\text{entropy}_{i\in\mathcal{B}}(\mathcal{I})}.
\eeqa
Both methods work by devising how well a method parses a region with a target from the rest (background). In Equations \eqref{eq:metrics}, $\mathcal{I}$ is the magnitude SAR image with entries $\mathcal{I}_i$ either in the target region $\mathcal{T}$ or background $\mathcal{B}$. TBR and TBED give a sense of how well a crafted SAR image is at distinguishing a target from the background and from errors that can plague such reconstructions (oscillations, blurring, etc). It has been shown that automatic target recognition schemes improve in performance with increasing TBED values for the training and testing image sets \cite{clark1991image,chen2008image}.

\subsection{Stripmap SAR Results}\label{subsec:bikeResults}

\begin{figure}\centering
\includegraphics[width=\columnwidth]{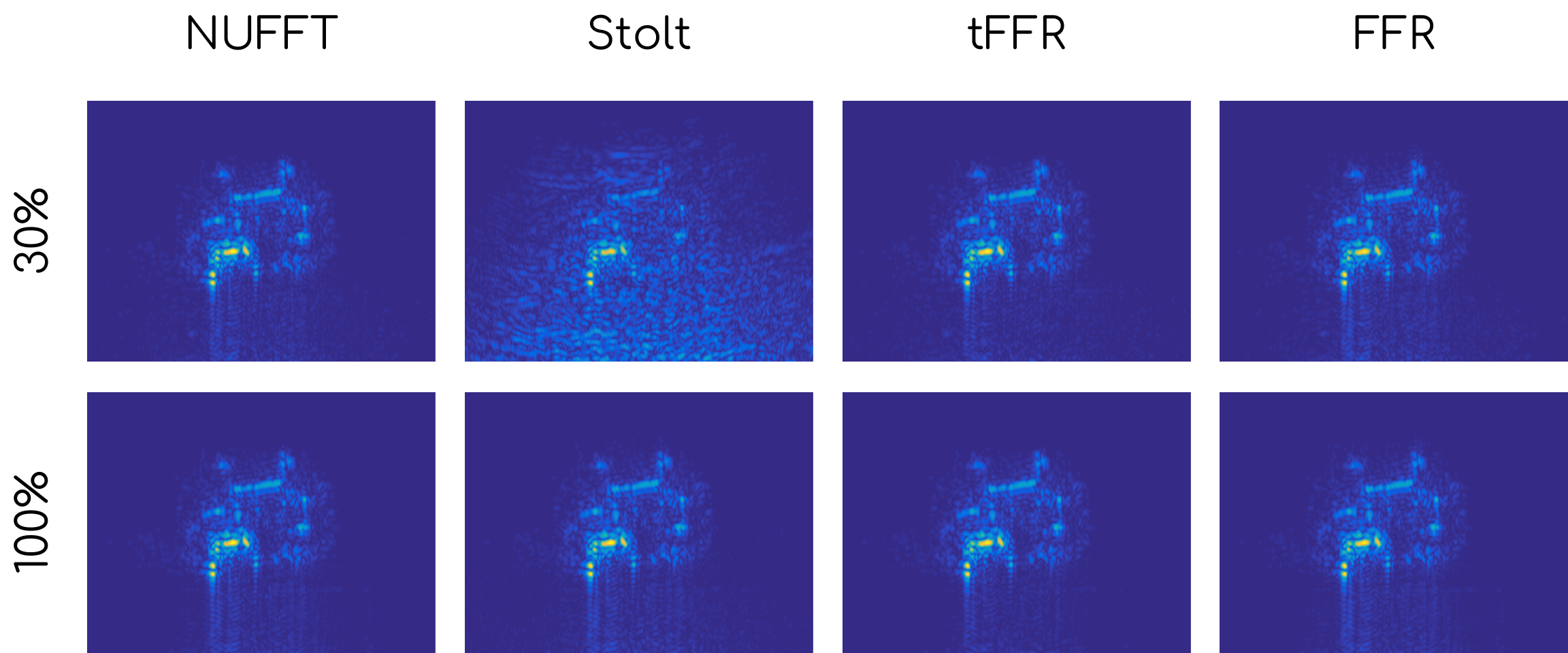}\\\vspace{.3cm}
\includegraphics[width=\columnwidth]{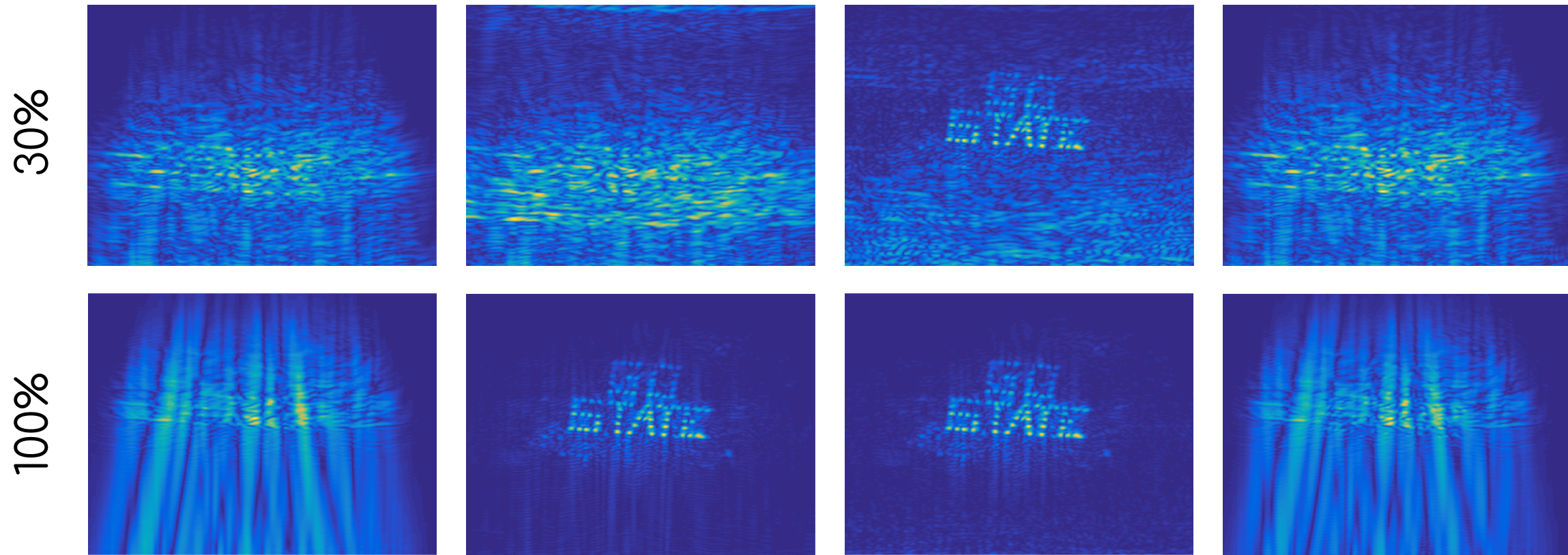}\\\vspace{.3cm}
\includegraphics[width=\columnwidth]{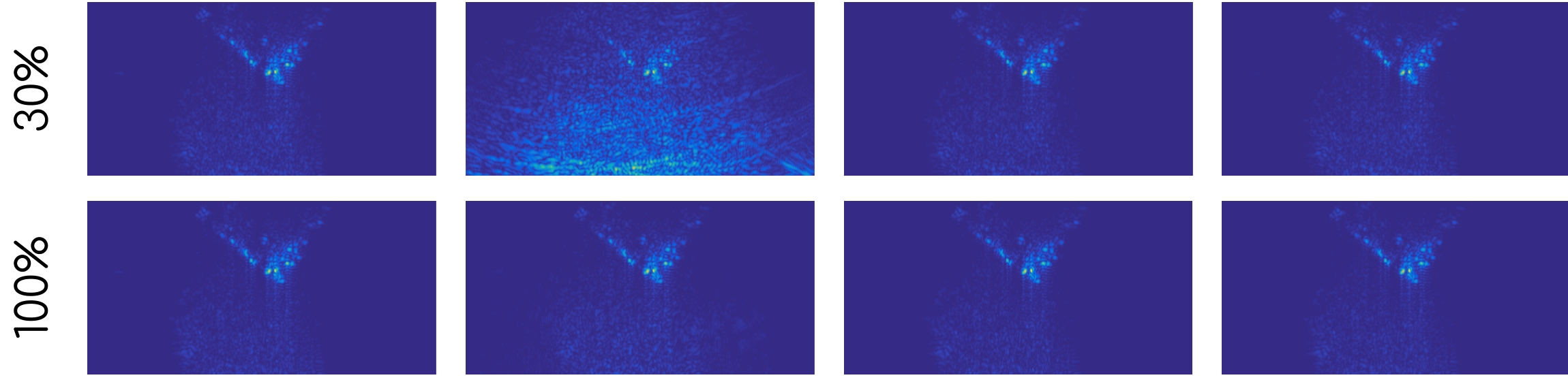}
\caption{Reconstruction results for the various methods for real stripmap SAR data, at lowest (30\%) and highest (100\%) sampling cases.}\label{fig:images}
\end{figure}

Overall we see that our tFFR method works best quantitatively in almost every scenario (See Figure~\ref{fig:tbr_tbed}). One reason for this success may be attributed to (i) the improved spatial localization in the radar images (validated by metrics), and (ii) the suppressed noise in the background. Figures~\ref{fig:images} show example SAR reconstructions from the lowest and highest sampled cases for each image. The ``Go State'' image turned out to be a failure case for the NUFFT and FFR methods due to their inability to adapt to this sampling configuration of data. Our tFFR reconstructs the image at higher TBED values than the Stolt, demonstrating the power of our method and its inherent flexibility. Further, we see that tFFR, based on both its reconstructed images and quantitative metrics, is able to perform well even in the severely limited sampling cases.

\begin{table}\centering\normalsize
\begin{tabular}{r|c|c|c|c|}
Image & Stolt & NUFFT & FFR & tFFR\\\hline
\textbf{Bike} & 1.01 & 25.7& 217.2 & 51.5 \\
\textbf{Go State} & .9 & 24.1 & 219.6& 70.01 \\
\textbf{Car} & .7 & 32.5 & 238.12& 86.7 
\end{tabular}
\caption{Computational time, in seconds, for reconstruction of the 100\% fully sampled images.}\label{tab:times}
\end{table}

As for computational time, we used the efficient \texttt{interp} function in \texttt{Matlab} for the Stolt interpolation as was done in \cite{charvat2006low}. This method was therefore implemented fastest, with approximately one second per image. The NUFFT and tFFR both were slower than Stolt,as reflected in Table \ref{tab:times}, which shows imaging time in seconds for the $100\%$ case (the most computationally expensive task). While our method tFFR requires between two and three times that the the NUFFT, the time is drastically reduced compared to the original FFR method which almost took around nine times as long as the NUFFT. We did not use any parallelization for our tFFR method and see this as a path towards an even faster implementation.

\section{Conclusion}
We detailed our contribution to the field of image recovery methods that combine the speed of NUFFTs with the sampling-robust characteristics of FAs. By employing a thresholding technique within the calculation for a correlation matrix, we were able to improve the computational expense of an existing but untested FFR algorithm with additional benefits including a lesser susceptibility to high frequency noise. We used real-world stripmap SAR data to demonstrate our tFFR method against FFRs, NUFFTs, and traditional Stolt interpolation techniques. We found that tFFR offers quantifiably better image reconstructions than those other known methods at a comparable computational expense.

\bibliographystyle{IEEEtran}
\bibliography{refGRSL2017}

\end{document}